\documentclass[10pt,aps,prb,superscriptaddress,twocolumn,showpacs,floatfix,longbibliography]{revtex4-2}
\usepackage{graphicx}
\usepackage{dcolumn}
\usepackage{bm}
\usepackage{amsmath}
\usepackage{subfigure}
\usepackage{url} 
\usepackage{hyperref} 
\usepackage{xcolor}

\usepackage{url}

\newcommand{\m}{\mathbf}
\newcommand{\h}{\hat}

\renewcommand{\t}{\theta}
\newcommand{\be}{\begin{eqnarray}}
\newcommand{\ee}{\end{eqnarray}}

\begin{document}


\title{Spin wave stiffness and damping in a frustrated chiral helimagnet Co$_8$Zn$_8$Mn$_4$ as measured by small-angle neutron scattering}

\author{V. Ukleev}
\email{victor.ukleev@psi.ch}
\thanks{V. Ukleev and K.A. Pschenichnyi contributed equally to this work.}
\affiliation{Laboratory for Neutron Scattering and Imaging (LNS), Paul Scherrer Institute (PSI), CH-5232 Villigen, Switzerland}
\affiliation{Swiss Light Source, Paul Scherrer Institute (PSI), CH-5232 Villigen, Switzerland}
\author{K.A. Pschenichnyi}
\affiliation{Petersburg Nuclear Physics Institute NRC “Kurchatov Institute”, Gatchina, St-Petersburg 188300, Russia}
\affiliation{Saint-Petersburg State University, Ulyanovskaya 1, Saint-Petersburg 198504, Russia}
\author{O. Utesov}
\affiliation{Saint-Petersburg State University, Ulyanovskaya 1, Saint-Petersburg 198504, Russia}
\affiliation{Petersburg Nuclear Physics Institute NRC “Kurchatov Institute”, Gatchina, St-Petersburg 188300, Russia}
\affiliation{National Research University Higher School of Economics, St.~Petersburg, 190008, Russia}
\author{K. Karube}
\affiliation{RIKEN Center for Emergent Matter Science (CEMS), Wako 351-0198, Japan}
\author{S. M\"uhlbauer}
\affiliation{Heinz Maier-Leibnitz Zentrum (MLZ), Technische Universit\"at M\"unchen, D-85748 Garching, Germany}
\author{R. Cubitt}
\affiliation{Institut Laue-Langevin, 71 Avenue des Martyrs, 38042 Grenoble, France}
\author{Y. Tokura}
\affiliation{RIKEN Center for Emergent Matter Science (CEMS), Wako 351-0198, Japan}
\affiliation{Department of Applied Physics, University of Tokyo, Tokyo 113-8656, Japan}
\affiliation{Tokyo College, University of Tokyo, Tokyo 113-8656, Japan}
\author{Y. Taguchi}
\affiliation{RIKEN Center for Emergent Matter Science (CEMS), Wako 351-0198, Japan}
\author{J. S. White}
\affiliation{Laboratory for Neutron Scattering and Imaging (LNS), Paul Scherrer Institute (PSI), CH-5232 Villigen, Switzerland}
\author{S. V. Grigoriev}
\affiliation{Petersburg Nuclear Physics Institute NRC “Kurchatov Institute”, Gatchina, St-Petersburg 188300, Russia}
\affiliation{Saint-Petersburg State University, Ulyanovskaya 1, Saint-Petersburg 198504, Russia}
\date{\today}

\newcommand{\czm}{Co$_8$Zn$_8$Mn$_4$}

\begin{abstract}
Multiple intriguing low temperature phenomena have recently been discovered in the family of chiral cubic Co-Zn-Mn compounds with $\beta-$Mn-type structure. In particular, \czm~displays a reduction of the helical spiral pitch on cooling, along with lattice shape transformations of metastable skyrmions and the manifestation of peculiar magnetic textures due to strong magnetocrystalline anisotropy. Here we report on temperature-dependent measurements of helimagnon excitations in the field polarized regime \czm~using the spin wave small-angle neutron scattering (SWSANS) technique. By applying a new analytical expression to interpret the data, quantitative estimates for both spin wave stiffness and damping are extracted across a wide temperature range between 70\,K and 250\,K. We speculate that their non-trivial temperature-dependencies arise due to the effects of magnetic frustration arising from Mn magnetic moments, which is further reflected in continuous variations of both exchange and Dzyaloshinskii-Moriya interactions.
\end{abstract}

\maketitle

\section{Introduction}

The competition between magnetic interactions such as the exchange interaction, Dzyaloshinskii-Moriya (DM) interaction, anisotropy and Zeeman energies is well known to result in omnifarious phase diagrams amongst noncentrosymmetric magnets \cite{dzyaloshinsky1958thermodynamic,moriya1960anisotropic}. Amongst the most studied are the chiral cubic magnets featuring zero-field helical, field-induced conical, skyrmion lattice (SkL) and fully polarized phases \cite{bak1980theory,nakanishi1980origin,roessler2006spontaneous}.

Recently, a new family of chiral cubic Co-Zn-Mn intermetallics with highly tunable magnetic properties was discovered \cite{tokunaga2015new}. These materials crystallize in the $\beta$-Mn-type structure with space group $P4_132$ or $P4_332$, with 20 atoms per unit cell distributed over two Wyckoff sites $8c$ and $12d$. The $8c$ site is mainly occupied by magnetic Co, while the $12d$ site is mainly occupied by nonmagnetic Zn or magnetic Mn \cite{tokunaga2015new,bocarsly2019deciphering,nakajima2019prb}. The magnetic properties of (Co$_{0.5}$Zn$_{0.5}$)$_{20-x}$Mn$_x$ solutions depend dramatically on the precise composition and exhibit complicated temperature-dependent magnetic phase diagrams \cite{karube2016robust,karube2017skyrmion,karube2018sciadv,karube2018controlling,karube2020metastable}. Notably, the helical spiral ordering temperature $T_C$ of 460\,K for the end member Co$_{10}$Zn$_{10}$ rapidly decreases with partial substitution of Mn, reaching $T_C = 300$\,K for \czm \cite{karube2020metastable}. Being important from the viewpoint of applications, this particular composition hosts an equilibrium SkL phase at room temperature under moderate magnetic fields of 50--100\,mT. Furthermore, in contrast to the archetypal $B$20-type chiral magnets, the spiral period $l_s$ in \czm~and other Co-Zn-Mn helimagnets (except Mn-free Co$_{10}$Zn$_{10}$) falls significantly on cooling below $T_C$. In \czm,~this leads to a large increase of the helical spiral vector $k_s=2\pi/l_s$ by $\sim$50~\%. The variation in $k_s$ is also reflected in the properties of the metastable SkL states, and contributes to the observed transition between a conventional triangular SkL coordination at high temperature, and either a square lattice or distorted array of L-shaped skyrmions below $T\sim100$\,K \cite{karube2016robust,morikawa2017deformation,ukleev2019element}.

\begin{figure*}[t]
\begin{center}
\includegraphics[width=17cm]{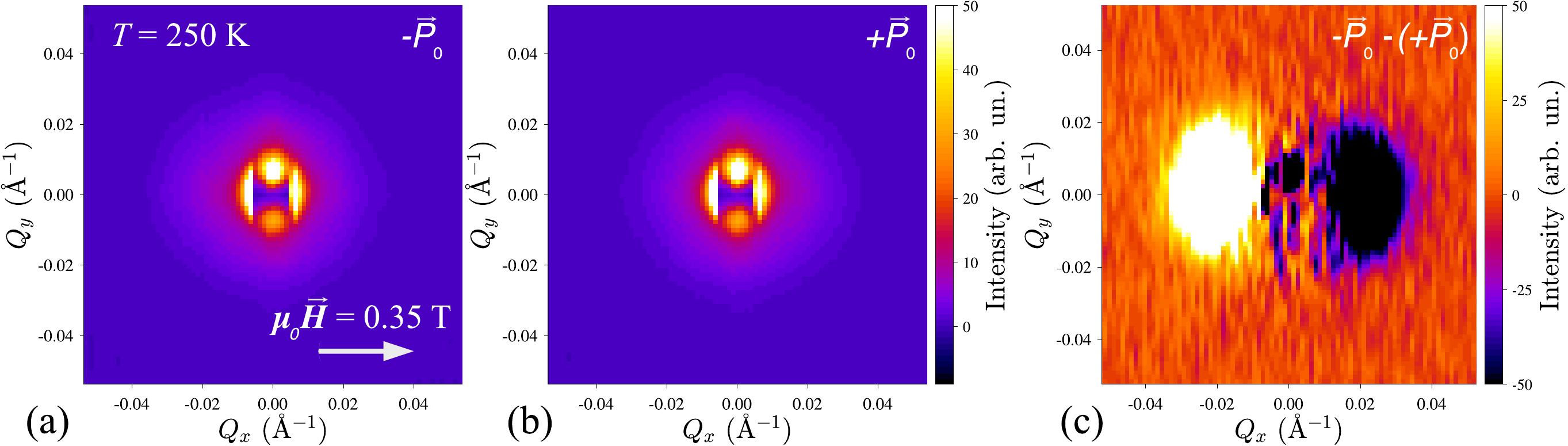}
\caption{Two-dimensional spin-polarized SANS maps measured at $T=250$\,K in the field-induced ferromagnetic state at applied magnetic field $\mu_0 H = 0.35$\,T with incoming neutron polarizations (a) $-\vec{P_0}$ and (b) $+\vec{P_0}$. The low-$Q$ region of the scattering patterns are dominated by the elastic backgound and direct beam contamination. c) Difference map of the $-\vec{P_0}$ and $+\vec{P_0}$ SANS patterns shown in panels (a,b).}
\label{fig1}
\end{center}
\end{figure*}

According to the Bak-Jensen model for chiral magnets, the characteristic helical wavevector $k_s=D/J$ depends on the Dzyaloshinskii constant $D$ and exchange integral $J$ \cite{bak1980theory,nakanishi1980origin}, providing a framework for connecting the unusual temperature variation of $k_s$ to the microscopic interactions. Beyond the clear correlation between this unusual behaviour and finite Mn content, a detailed study of the connection between the observation and the fundamental interactions in Co-Zn-Mn compounds has not been done yet. From micromagnetic simulations, both the variation in $k_s$ and corresponding deformation of SkLs in a \czm~thin plate can be reproduced by incorporating a linear decrease (increase) of $J$ ($D$) in the presence of cubic anisotropy that mimics the effect of decreasing $T$ \cite{ukleev2019element}. To gain deeper insight however requires direct measurement of the microscopic magnetic interactions, without mutual dependence on one another through the measurement of parameters such as $k_s$. In the present work, we perform measurements of the spin-wave (SW) stiffness of \czm~which provides direct measurements of $J$ that are decoupled from the DM interaction parameter $D$. 

While it is well known that the fingerprint of microscopic interactions are directly manifested in the SW dispersion relations, it is not straightforward to apply standard inelastic neutron scattering, microwave, and optical spectroscopy tools to Co-Zn-Mn compounds due to kinematic limitations, the limited volume of the single-crystalline samples and large SW damping in metallic alloys. Instead, we apply the new spin-wave small-angle neutron scattering (SWSANS) method that allows a direct determination of spin-wave stiffness in the field-polarized state of helical magnets. The technique has been proposed earlier and utilized successfully for $B$20-type compounds \cite{grigoriev2015spin,siegfried2017spin,grigoriev2018spinFeGe,grigoriev2018spinMnFeSi,pshenichnyi2018calculation} and Cu$_2$OSeO$_3$ \cite{grigoriev2019spinCu2O}. All of these compounds display modest reductions of $k_s$ as $T$ falls below $T_C$, and it was found by SWSANS that the spin wave stiffness parameter $A_{\textrm{ex}}$ displays some softening near $T_C$ before increasing towards low $T$. From the present SWSANS study of \czm~we demonstrate that the $T$-variation of $k_s$ cannot be explained solely by a $T$-variation of the exchange integral, and we draw conclusions concerning the hybrid nature of the helical pitch shortening in \czm~at low $T$ that involves variations in both exchange and DM interactions. Further, we evidence the strong damping of spin waves below $T\approx 150$\,K which is likely caused by frustration-induced static and dynamic disorder of antiferromagnetically-interacting Mn spins.

\section{Experimental}

The SANS measurements were performed using the SANS-1 instrument \cite{muhlbauer2016new} at the Heinz Maier-Leibnitz Zentrum (MLZ, Garching, Germany) and the D33 instrument \cite{dewhurst2016small} at the Institut Laue-Langevin (ILL, Grenoble, France). At both instruments, a mean neutron wavelength $\lambda = 5$\,\AA~was chosen and the beam was collimated over 8\,m before the sample. The scattered neutrons were counted by a two-dimensional position-sensitive detector located 8\,m behind the sample. The crystal was mounted on aluminum holder and installed in a horizontal field cryomagnet (5\,T and 2.5\,T at SANS-1 and D33, respectively). For the polarized neutron scattering experiment at SANS-1 instrument a neutron beam with initial polarization $P_0 = 0.93$ was produced by V-shaped FeSi transmission polarizer. The sample was the same single crystal of \czm~characterized previously by magnetometry and elastic SANS \cite{karube2016robust}.

\section{Results and Discussions}

\begin{figure*}[t]
\begin{center}
\includegraphics[width=17cm]{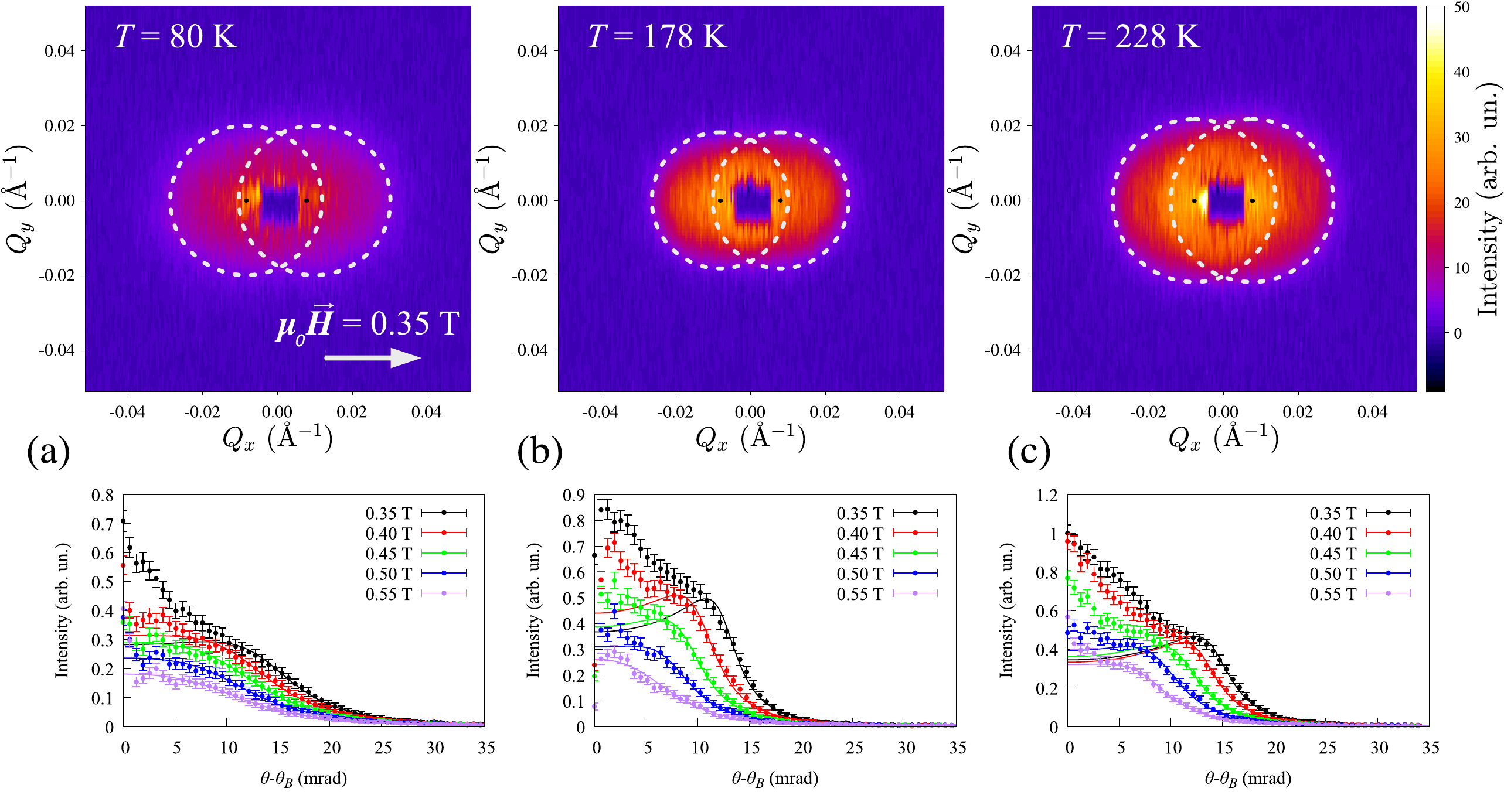}
\caption{SWSANS maps (top panel) and azimuthally averaged intensities  (bottom panel) measured at (a) $T=80$\,K, (b) $T=178$\,K and (c) $T=228$\,K at $\mu_0 H = 0.35$\,T. The origin of SWSANS diffuse spots at $Q = \pm k_s$ is marked by a black dot. The solid lines in the intensity $I(H,\,\theta-\theta_B)$ plots for various magnetic fields represent the best fit using Eq. \ref{newfit}.}
\label{fig2}
\end{center}
\end{figure*}

The energy carried by a spin-wave in a field-polarized helimagnet is given by Kataoka as \cite{kataoka1987spin}:
\begin{equation}
\epsilon(\mathbf{Q})=A_{\textrm{ex}}(\mathbf{Q}-\mathbf{k_s})^2+(H-H_{c2}),
\label{energy}
\end{equation}
where $H_{c2}$ is the upper critical field, and the propagation vector $\mathbf{k_s}$ matches the orientation of the applied magnetic field $H$. The direction of nonreciprocal propagation is either parallel or antiparallel to $H$ and determined by the sign of the DM constant $D$ \cite{iguchi2015nonreciprocal,sato2016magnon,seki2016magnetochiral}. Correspondingly, the SANS cross sections for SW scattering are antisymmetric for incoming neutron spin polarizations $\pm \vec{P_0} \approx 0.93$ either aligned or antialigned with $H$. This was first demonstrated for MnSi in Ref.\cite{grigoriev2015spin} and confirmed in present work for the \czm~single crystal.

Figure \ref{fig1} shows polarized SANS data measured from \czm~at $T=250$\,K after zero field cooling. A magnetic field of $\mu_0 H=0.35$\,T was applied along the [100] axis antiparallel to the neutron spin polarization $\vec{P_0}$ and transverse to the incoming neutron beam. The applied magnetic field was larger than $\mu_0 H_{c2}=0.25$\,T \cite{karube2016robust} and sufficient to saturate the sample and remove the contribution to the SANS signal due to elastic scattering from the conical state. The remaining signal at low $Q$ represents the $Q=0$ scattering from ferromagnetic correlations \cite{altynbaev2014intrinsic,white2018direct} and some remaining background arising from the direct beam tail and imperfect background subtraction.
The inelastic SW scattering appears as a broad diffusive spot feature centered at $Q = \pm k_s$ (Figs. \ref{fig1}a,b). Figure~\ref{fig1}c shows a difference image of SANS intensities measured with opposite neutron polarizations $\pm \vec{P_0}$ which more clearly signifies the relative difference in scattering due to nonreciprocal SW propagation, and is similar to that observed in MnSi \cite{grigoriev2015spin}. The radius of a diffuse scattering spot is limited by a critical scattering angle $\theta_c$ that depends on $H$. According to Eq.~\ref{energy}, the SW dispersion narrows with increase of the field and disappears at a certain $\mu_0 H_\textrm{off}$. We thus define the cutoff angle as
\be
\theta_c^2(H)=\theta_0^2 - \theta_0 g\mu_B (H-H_{c2})/E_i,
\label{thetac}
\ee
where $\theta_0 = \hbar^2 / (2 A_{\textrm{ex}} m_n)$, $m_n$ is the neutron mass, and $E_i$ is the energy of incident neutrons. Eq.~\ref{thetac} allows the determination of the spin-wave stiffness parameter $A_{\textrm{ex}}$ from measurements of the $H$-dependence of the cutoff angle $\theta_c(H)$. The spin-wave stiffness is thus defined as a solution to Eq.~\ref{thetac}
\be
\begin{gathered}
    \label{eq:stiffness}
    A_{\textrm{ex}} = \frac{\hbar^2}{2m_nE_i} \
    \{\frac{E_i^2}{\sqrt{E_i^2\theta_{C}^2+g^2\mu_B^2(H-H_{c2})^2}} + \\ + g\mu_B(H-H_{c2})\},
\end{gathered}
\ee
averaged over several magnetic fields. From $A_{\textrm{ex}}$, the exchange integral $J$ can be determined through the relation $A_{\textrm{ex}} = SJ$ is the spin-wave stiffness and $S$ is the ordered spin. 

Having proved the inelastic SW origin of the SANS scattering in the field-polarized regime using polarized neutrons, the $T$-dependence of the SWSANS signal was measured using unpolarized neutrons, which allowed the measurements at each $T$ to be done in a comparatively shorter time. $T$-dependent measurement were done on warming after an initial ZFC from $T>T_C$ to 15\,K. Before starting the $T$-dependent measurements, a background measurement was done at 15\,K under a high field $\mu_0 H = 3$\,T that was sufficient to fully suppress the observable SW signal. Subsequently, at each $T$ the $H$-dependence of the elastic SANS signal in the helical and conical phases was measured to experimentally determine the value $k_s$. Finally, SWSANS patterns were measured as a function of magnetic field $H>H_c$ in field-polarized phase. In the top half of Figs.\ref{fig2}(a)-(c), we show background corrected SANS patterns of the SWSANS intensities measured at $\mu_{0}H$=~0.35\,T at different temperatures. Bottom panels of Figs.\ref{fig2} show intensity $I(H,\,\theta-\theta_B)$ plots for various magnetic fields, where $\theta_B$ corresponds to the Bragg angle of helical peak. To improve the statistics, the scattering intensity of the SANS maps was azimuthally averaged over the angular sector of $120^\circ$.


In previous SWSANS studies, the cutoff feature of the azimuthally-averaged intensity $I$ was treated using a phenomenological step-like {\it arctan} function \cite{grigoriev2015spin,siegfried2017spin,grigoriev2018spinFeGe,grigoriev2018spinMnFeSi,pshenichnyi2018calculation,grigoriev2019spinCu2O}. In the present study on \czm, the analogous cutoff feature is smeared due strong magnon damping, which is an aspect that needs to be properly accounted for in the data analysis in order to obtain reliable estimates of $A_{\textrm{ex}}$. Here we propose a different function which is justified theoretically (see the Appendix) and which provides an excellent fit of the data, both in the vicinity of the cutoff angle and at larger scattering angles
\be
  I(\t) = I_0 \, \textrm{Im} \frac{1}{\sqrt{(\t-\t_B)^2 -\t^2_C - i \gamma_{0}}}.
  \label{newfit}
\ee
Here $I_0$ is a fitting parameter, and $\gamma_0$ is the dimensionless damping parameter ($\gamma_0 = \t_0 \Gamma/E_i$) for magnons with energy $2 \t_0 E_i$. This equation is derived under an assumption, that $\gamma_0 \ll \t^2_C$. When the magnon damping is large (or $\t_C$ is small) $\gamma_0 \gtrsim \t^2_C$ and Eq.~\eqref{newfit} can not be applied. At the same time, the concept of a cutoff angle also becomes questionable since magnons can no longer be considered as well-defined quasiparticles. Bearing these considerations in mind, in our analysis we found that we were unable to treat reliably the experimental SWSANS data in the lower temperature range $T<70$\,K.

\begin{figure}
\includegraphics[width=8.5cm]{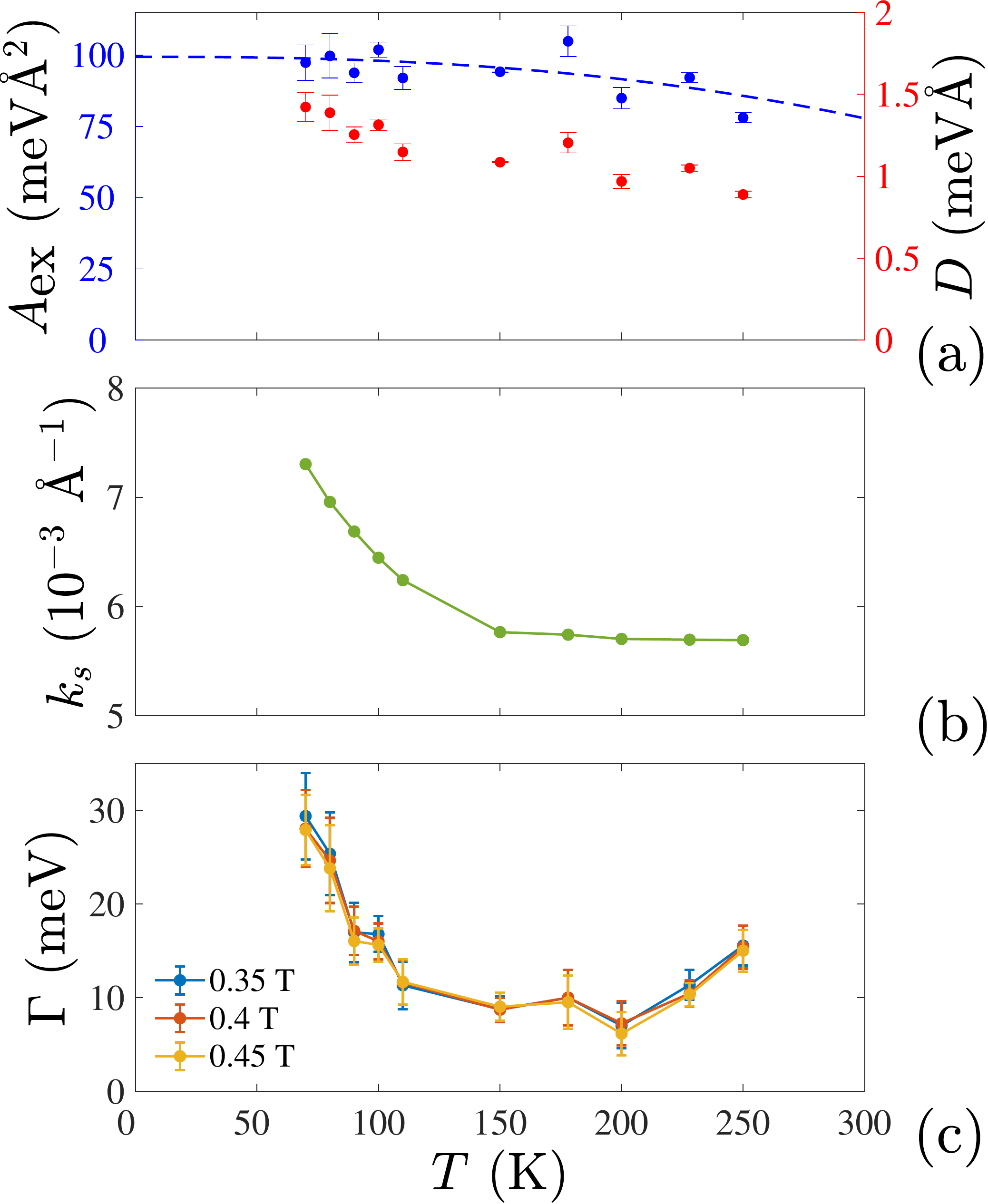}
\caption{(a) The temperature dependence of the spin wave stiffness $A_\textrm{ex}$ (blue symbols, left axis) and DMI $D$ (red symbols, right axis) obtained from the cutoff angles measured at different applied magnetic fields according to the Eq. \ref{thetac}, and from the temperature dependence of $A_{\textrm{ex}}$ and $k_s$ (shown in the next panel). The dashed line is the $A_\textrm{ex}(T)$ fit according to the power law. (b) The temperature dependence of the spiral wavevector $k_s$. (c) Spin wave damping $\Gamma$ variation as a function of temperature obtained from the Eq. \ref{newfit} at each temperature and magnetic fields of 0.35\,T, 0.4\,T and 0.45\,T.}
\label{fig3}
\end{figure}

The main findings of the present work obtained from the SWSANS data are shown in Figs.~\ref{fig3}(a) and (b). Displaying a similar behaviour as previously determined for the $B$20-type materials, Fig.~\ref{fig3}(a) shows $A_{\textrm{ex}}(T)$ tends to increase on cooling from 250\,K to 70\,K. We also anticipate that $A_{\textrm{ex}}(T)$ in \czm~remains finite at $T_C = 300$\,K analogous to $B$20s and Cu$_2$OSeO$_3$. In order to extrapolate the measured values of the spin wave stiffness to $T = 0$, the temperature dependence $A_{\textrm{ex}}(T)$ was fitted according to the power law: $A_{\textrm{ex}}(T)=A_0(1-c(T/T_C)^z)$, where $z = 5/2$ is the fixed power law as can be expected for the ferromagnets \cite{mezei1984critical}, and the fitted parameters are $A_0=99\pm6$\,meV\AA$^2$ and $c=0.22\pm0.05$. By using the values of $k_s$ measured by elastic SANS in zero field Fig.~\ref{fig3}(b), and the simple Bak-Jensen relation $k_s = D/J$, the $T$-dependence of the effective DM interaction is also obtained and shown in Fig.~\ref{fig3}(a). 

It becomes clear that within this framework, the large increase of $k_s$ by $\sim50\%$ below 120\,K (Fig.~\ref{fig3}(a); more details on temperature dependence of $k_s$ are given in Ref. \onlinecite{karube2016robust}) cannot be explained solely by a variation of exchange integral $J$, and thus the DM parameter $D$ is also required to increase at low $T$s (red symbols in Fig.~\ref{fig3}(a)). Indeed, a simple relation between the exchange stiffness and the critical field $H_{c2}$ in Bak-Jensen model $g \mu_B H_{c2}=A_{\textrm{ex}} k_s^2$ implies a significant downturn of $A_{\textrm{ex}}(T)$ below 120\,K, which is not observed experimentally. At the same time at low $T$s, we determine a dramatic increase of the damping parameter $\Gamma$ as shown in Fig.~\ref{fig3}(c), which contrasts strongly to the previously studied $B$20 systems. While it is unfortunate that the strong damping imposes a limit on the applicability of Eq.~\ref{newfit} so that the microscopic parameters cannot be determined for $T<70$\,K, we note that this physically is consistent with the observed onset of magnetic disordering on the approach to a spin glass regime below $\sim$10\,K \cite{karube2020metastable}. At high temperatures the damping parameter diverges when the system approaches the critical temperature, which is consistent with previous reports on $B$20 compounds \cite{siegfried2017spin,grigoriev2018spinMnFeSi,grigoriev2018spinFeGe,grigoriev2019spinFeCoSi,grigoriev2019spinCu2O}.

Here we suggest a few possible origins of the observed $T$-dependence of the DM parameter. Firstly, an increase in $D$ can be explained by the monotonic increase of the magnetic moment and the corresponding hybridization of the orbitals near the Fermi level which essentially determines the size of the DMI in metals \cite{koretsune2015control,gayles2015dzyaloshinskii}. A similar effect has been recently observed in 3$d$-5$d$ metallic multilayers with interfacial DMI \cite{schlotter2018temperature,zhou2020temperature}. However, this mechanism does not explain the enormous increase of SW damping in \czm~below 150\,K.

Therefore, turning to both the DM and SW damping parameters, we attribute their observed $T$-dependences to a corresponding $T$-evolution of magnetic frustration of the antiferromagnetically interacting Mn. At low $T$ the net magnetization due to Mn moments effectively vanishes compared to the contribution of Co \cite{ukleev2019element,ukleev2021frustration}, while \textit{ab-initio} calculations suggest an opposite scenario with $\mu_\textrm{Co}\approx1.3$\,$\mu_\textrm{B}$ and $\mu_\textrm{Mn}\approx3.2$\,$\mu_\textrm{B}$ (or 2.4\,$\mu_\textrm{B}$ if it is on its minority $8c$ site) \cite{bocarsly2019deciphering}. The frustration results in the experimentally observed short-range magnetic correlations and slow magnetic fluctuations induced by the antiferromagnetic exchange between Mn moments on the 12$d$ site \cite{ukleev2019element}. The magnetic disorder introduced by antiferromagnetic Mn-Mn correlations prevents the coherent propagation of helimagnons and results in the large SW damping. Indeed, the low-temperature damping enhancement that was observed even in the field-induced ferromagnetic phase in the present neutron and in the previous ferromagnetic resonance spectroscopy \cite{preissinger2021vital} experiments suggests the antiferromagnetic nature of the effect that cannot be easily suppressed by magnetic fields of 1\,T or below. Further, a recent theoretical work suggests that the frustration of Heisenberg spins in a chiral magnet results in a rich variety of the possible spiral ground states with the period determined by the competition between ferromagnetic and antiferromagnetic exchanges, and DMI \cite{mutter2019skyrmion}. We suppose, that in our case the effective $D$ parameter is determined by both antisymmetric DMI and frustrated exchanges and shows the non-trivial temperature dependence due to their interplay.

Another further contribution to anomalous variation of the helical pitch can in principle arise from the anisotropic exchange interaction (AEI), which is often ignored due to its weak contribution in, for example, the $B$20-type magnets. Within the Bak-Jensen model the expression for the helical propagation vector that takes the AEI into account reads as \cite{maleyev2006cubic}
\begin{equation}
k_s=\frac{D}{2J} (1 - \frac{F}{4 J} L(\hat{k})), 
\label{aex}
\end{equation}
where $F$ is anisotropy constant and $L(\hat{k}) = 2 \sin^2 \psi (\sin^2 \phi \cos^2 \phi + \cos^2 \psi)$ is a cubic invariant determining the orientation of $k_s$ relative to the crystal axes with the corresponding angles $\psi$ and $\phi$. The cubic invariant describes the energy landscape when $L(\hat{k})$ is minimal and equal to 0 when $\hat{k}$ is oriented along the principal cube axes, and maximal when $\hat{k}$ is parallel to cubic diagonals. The AEI favors ground-state spirals propagating along $(100)$-equivalent axes in case of $F<0$ (Cu$_2$OSeO$_3$, Fe$_{0.85}$Co$_{0.15}$Si, Co-Zn-Mn) and  $\hat{k}||(111)$ if $F>0$ (MnSi). We point out that a strong increase in the AEI constant towards low $T$ could also result in an observable increase in magnitude of $k_s$. The possibility for this scenario is hinted from the $T$-dependence of the helical-to-conical transition field \cite{karube2016robust}, and also from the increase of the magnetocrystalline anisotropy at low $T$ evidenced by ferromagnetic resonance \cite{preissinger2021vital}. This mechanism of the spiral pitch variation with temperature does not imply the essential role of the DMI. However, it requires further investigation with the information of magnetocrystalline and exchange anisotropies unambiguously discriminated. 

It is likely, that all three above mentioned scenarios of the spiral pitch variation, namely: 1) temperature-dependent spin-orbit coupling; 2) interplay between the frustrated exchanges and DMI; and 3) emergence of the strong AEI, are relevant in case of the Co-Zn-Mn family of materials due to their structural and magnetic complexities. Experimentally, the nearest-neighbor and further-neighbor exchange parameters should be determined in further inelastic neutron scattering experiments, and the impact of the AEI on the helical pitch should be addressed and quantified by further high-resolution small-angle neutron or resonant x-ray scattering on single crystal samples \cite{ukleev2021signature,moody2021experimental}.

\section{Conclusions}

In conclusion, we have observed a non-trivial $T$-dependence of the spin wave stiffness $A_{\rm{ex}}$ in the \czm~chiral magnet by spin-wave small-angle neutron scattering (SWSANS). In contrast to the previously studied $B$20-type magnets, the spin-wave damping significantly influences the SWSANS signal, and a proper account of it is required for interpreting the data and the quantitative extraction of microscopic parameters from the data. Therefore, we proposed a new theoretical model for the SWSANS intensity, which accounts more rigorously for the damping effect and can be further applied to extract more details from the spin-wave spectra of other chiral helimagnets. Moreover, we further determine a significant increase in DMI at low temperatures which contributes to the observed reduction of the helical spiral pitch on cooling. Alternatively, the variation in spiral pitch might also be explained by an interplay of DMI and frustrated exchanges, or an increase in strength of anisotropic exchange. All these mechanisms are predicted to result in rich variety of the theoretically predicted spin textures that cannot be expected due to exchange and Dzyaloshinskii-Moriya interactions alone \cite{leonov2015multiply,mutter2019skyrmion,qian2018new,chacon2018observation,bannenberg2019multiple}. Indeed, multiple exotic real-space magnetic patterns, such as square skyrmion lattice \cite{karube2016robust}, L-shaped skyrmions \cite{morikawa2017deformation,ukleev2019element}, meron-antimeron lattice, disordered skyrmions \cite{yu2018transformation}, smectic liquid-crystalline structure of skyrmions \cite{nagase2019smectic} and domain-wall bimeron chains \cite{nagase2020observation} have been observed in the rich family of Co-Zn-Mn compounds. We believe that by combining the determination of the underlying magnetic interaction parameters with further theoretical and experimental studies, deeper insight can be obtained on the origin and stability of such spin textures, and inspire new approaches for their controllable creation, annihilation and manipulation in spintronics devices. 

\section*{Data availability}

All experimental data presented in the figures that support the findings of this study are available from the \href{https://doi.org/10.5281/zenodo.6572950}{Zenodo repository}\cite{zenodo2022}. Raw SWSANS data from D33 is also available from the ILL repository \cite{illdata}.

\section*{Acknowledgements}

The authors thank T. Arima, T. Nakajima, R. Takagi, and H.M. R{\o}nnow for helpful discussions. This research was supported in part by  JST CREST (Grant No. JPMJCR20T1 and JPMJCR1874). V.U. and J.S.W. acknowledge funding from the SNSF Projects No. 200021\_188707 and Sinergia CRSII5\_171003 NanoSkyrmionics. V.U. acknowledges financial support from the SNSF National Center of Competence in Research, Molecular Ultrafast Science and Technology (NCCR MUST). The authors thank the Heinz Maier-Leibnitz Zentrum and Institut Laue-Langevin for provision of neutron beamtime according to the proposals No. 15362 and 5-42-535 \cite{illdata}, respectively. The contribution to the study by O.U. was funded by the Russian Federation President Grant No. MK-1366.2021.1.2.

\appendix

\section{Appendix: SWSANS intensity and magnon damping}

Here we discuss the theoretical basis behind Eq.~\eqref{newfit}. In the limiting case of zero damping addressed in Ref.~\cite{grigoriev2015spin}, the inelastic magnetic chiral contribution to the neutron cross section $\sigma_{ch}(\m Q, \omega)$ can be represented by~\cite{maleev2002polarized}
\begin{align}
  \sigma_{ch}(\m Q, \omega) = \frac{k_f}{k_i}2r^2 |F_m|^2 \frac{1}{\pi\left(1-e^{-\omega/T}\right)} \left\langle S \right\rangle P_0 (\m {\h Q} \m{\h h})^2 \nonumber\\ \times \left[\delta(\omega-\epsilon_{\m Q}) + \delta(\omega+\epsilon_{-\m Q})\right],
  \label{A1}
\end{align}
where $k_f$ and $k_i$ are momenta of the scattered and incident neutron, respectively, $r$ is the classical electron radius, $F_m$ is the magnetic form factor, $\left\langle S \right\rangle$ is the average atomic spin, $\m {\h Q}$ is the unit vector along the momentum transfer, and $\epsilon_{\m Q}$ represents the spin wave dispersion. This equation assumes that the initial polarization is directed along the unit vector of an applied magnetic field: $\m P_0 = P_0 \m{\h h}$, $\m{\h h} = \m H/H$. Importantly, all which follows is also valid for unpolarized neutrons when the cross-section is symmetrical with respect to $\m{Q} \rightarrow -\m{Q}$ (it differs strictly from~\eqref{A1} only by the signs between delta functions and the absence of the $P_0$ factor).

In a Cartesian basis, we define the $z$-axis along the incident beam and the $x$-axis along the applied magnetic field. Therefore, $Q = k_i(\t_x,\t_y,\omega/2E_i)$ if $\omega \ll E_i$. Under the condition $\omega \ll T$, one can replace  $\left(1-\exp(-\omega/T)\right)^{-1}$ in Eq.~\eqref{A1} by $T/\omega$. This formula should be integrated over $\omega$ for consistency with the SANS experiments. For further theoretical discussion we introduce a dimensionless transferred energy $t = \omega/2 E_i$ and obtain the cross-section in the following form:
\begin{align}
  \sigma_{ch}(\t) \sim  \left\langle S \right\rangle T P_0 \int  \frac{d t}{t} \frac{\t^2_x}{t^2 + (\t_x^2 + \t_y^2)}  \nonumber\\  \times\left[\delta(t-\epsilon_{\m Q}/2E_i) + \delta(t+\epsilon_{-\m Q}/2E_i)\right].
  \label{Int1}
\end{align}
Delta-functions here represent the energy conservation law. Their sum can be rewritten as
\be
   \frac{\delta(t-t_1)+\delta(t-t_2)}{|1-t/\t_0|}+\frac{\delta(t-t_3)+\delta(t-t_4)}{|1+t/\t_0|}, 
\ee
where 
\be
  t_{1,2} = \t_0 \pm \sqrt{\t_C^2 - (\t_x-\t_B)^2 - \t_y^2}, \\
  t_{3,4}= - \t_0 \pm \sqrt{\t_C^2 - (\t_x+\t_B)^2 - \t_y^2}. \nonumber
\label{A2}
\ee
Here we use dimensionless parameters $\t_B, \t_C, t_C$ which are introduced after Eq.~\eqref{thetac}. Note, that $\t_C \leq \t_0$.

\begin{figure*}[t]
\begin{center}
  \includegraphics[width=5.7cm]{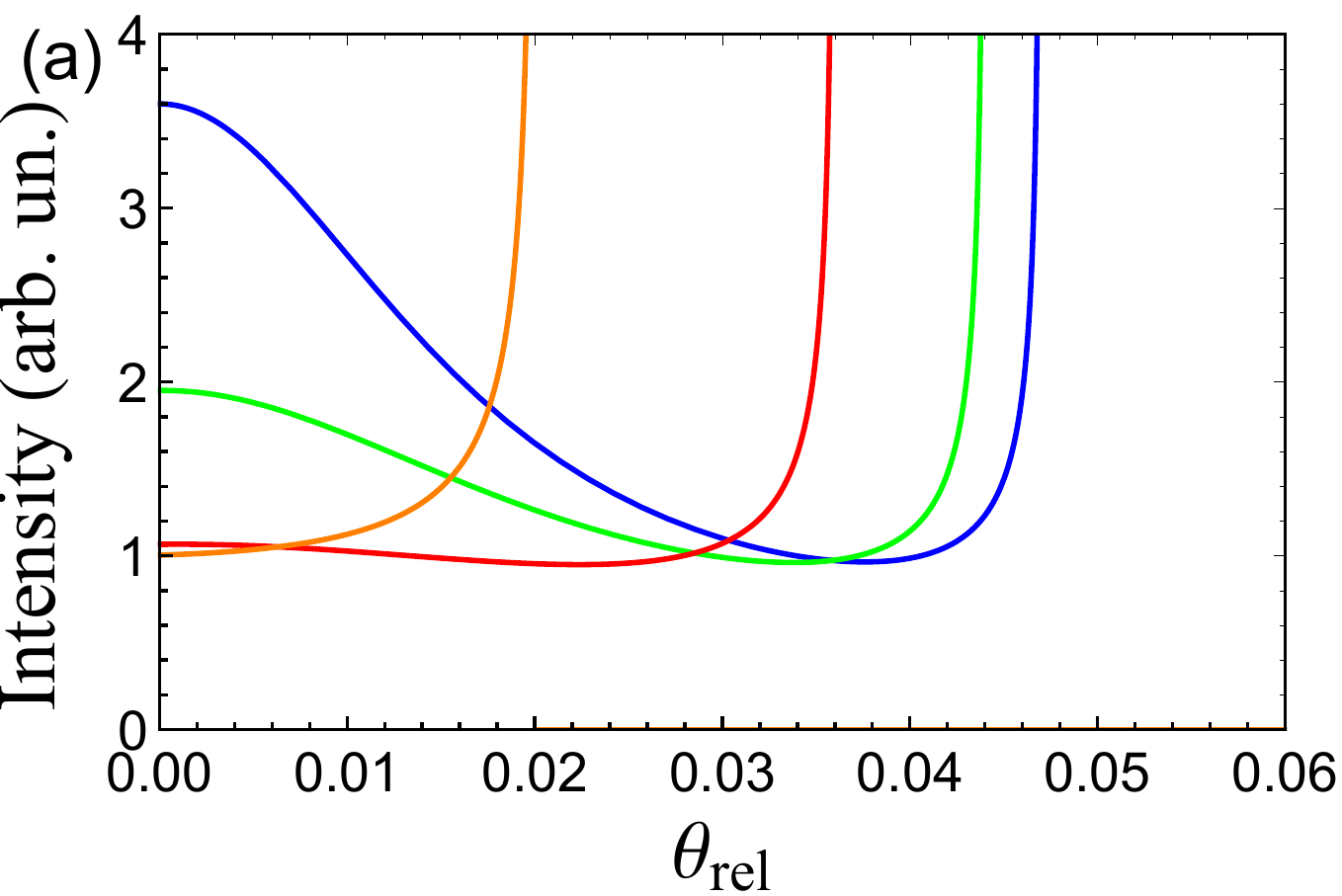}
  \includegraphics[width=5.7cm]{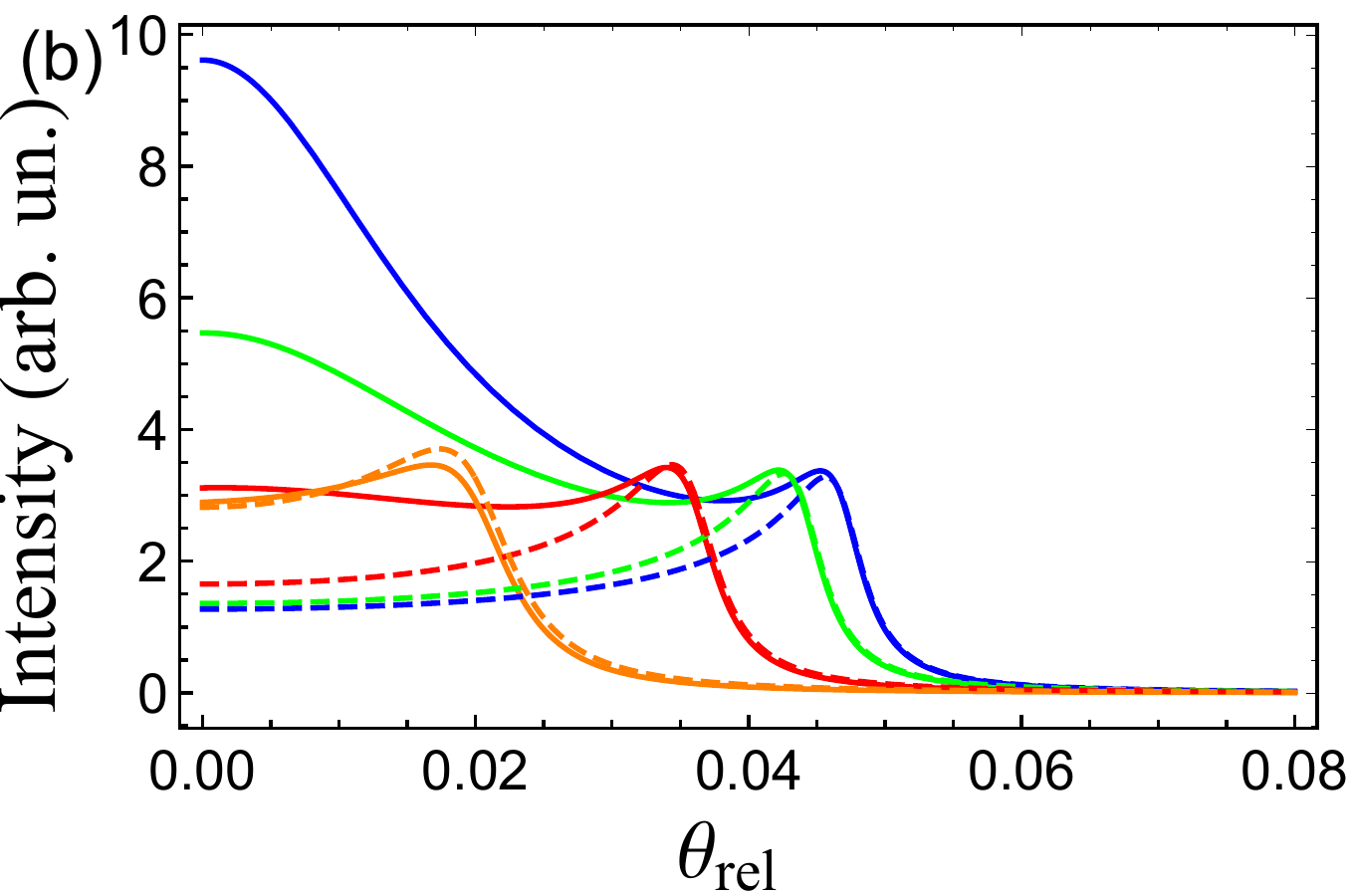}
  \includegraphics[width=5.7cm]{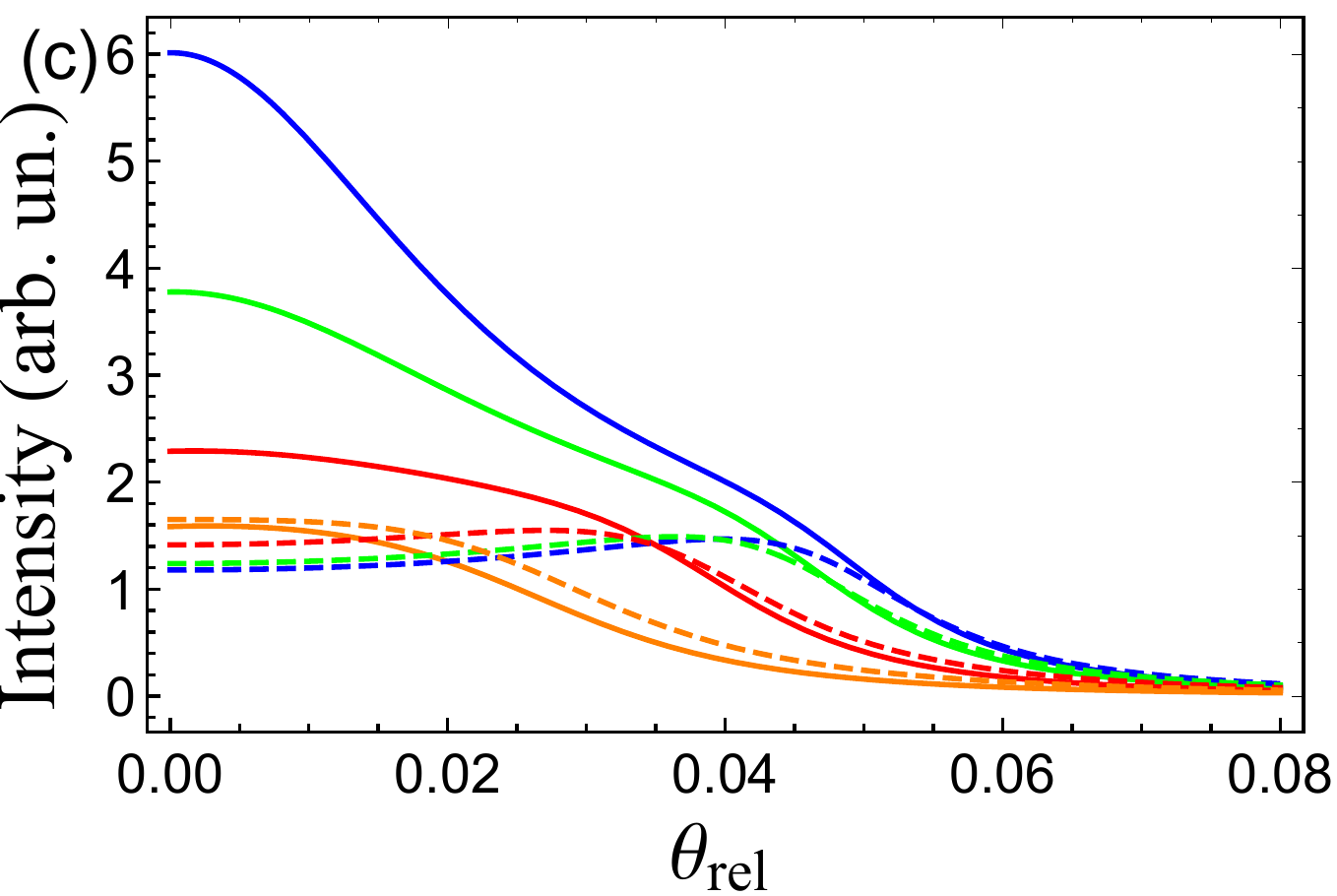}
  \caption{Sketch of the scattering intensity as a function of $\t_{rel}$ (distance from the Bragg angle) in different regimes. $\t_B=0.1, \t_0 = 0.05$ everywhere. The blue curves are for $\t_C = 0.047$, the green curves are for $\t_C = 0.044$, the red curves are for $\t_C = 0.036$, and the orange ones are for $\t_C = 0.02$. (a) SWSANS spectra without magnon damping. One can see the divergence at the cutoff angle, cf. Eq.~\eqref{Asigma1}. (b) SWSANS spectra for relatively small magnon damping $\gamma_0 \approx 0.0002$. Solid curves here stand for the spectrum calculated numerically using Eq.~\eqref{Int1} with the substitutions~\eqref{subs}, whereas dashed curves are plotted using analytical formula~\eqref{Asigma2}. (c) The same as (b), but for larger damping parameter $\gamma_0 \approx 0.001$. One can see that the peculiarity at $\t \approx \t_C$ is smeared out.
   }\label{FigApp}
\end{center} 
\end{figure*}

Further, we consider a standard situation where $\t_B > \t_0$ and where the scattering spheres for $\t_x>0$ and $\t_x<0$ do not overlap. It is also convenient to define a relative angle $\t_{rel}= \sqrt{(\t_x - \t_B)^2 + \t^2_y}$. The signal for $\t_x>0$ is determined only by the solutions $t_{1,2}$. The minimal and maximal $t = \t_0 \pm \t_C$ corresponds to $\t_x=\t_B, \, \t_y =0$ ($\t_{rel}=0$), whereas at the cutoff $\t = \t_C(H)$ we have equal solutions $t_{1,2} = \t_0$. The solution $t = \t_0 - \t_C$ results in significant growth of the scattering intensity near the Bragg angle for $\t_C \approx \t_0$ (due to the $1/t$ factor in the integral~\eqref{Int1}), while the degeneracy at the cutoff provides a hyperbolic increase of the intensity near $\t_C(H)$. Indeed, we have $t \approx \t_0 \pm \sqrt{2\t_C(H) (\t_C(H) - \t_{rel})} $ for $\t_{rel} \approx \t_C(H)$, thus
\be
  \sigma_{ch}(\t) \propto 1/\sqrt{\t_C(H) - \t_{rel}}.
  \label{Asigma1}
\ee
Experimentally this divergence is suppressed by instrumental resolution and magnon damping (see below). The SWSANS spectrum without damping is thus shown in Fig.~\ref{FigApp}a).


Next, we discuss how to properly introduce the physical effect of magnon damping into the equations above. In the fully polarized state, correlation functions of transverse spin components in momentum space are related to the simple magnon Green's function
\be
  G(\omega, \m{Q}) = \frac{1}{\omega - \epsilon_{\m{Q}} + i \delta},
  \label{G1}
\ee
where $\delta \rightarrow +0$ shows retarded character of the Green's function. In the presence of disorder (and/or other magnon scattering mechanisms, e.g. due to interaction with phonons or free electrons) one has
\be
  G(\omega, \m{Q}) = \frac{1}{\omega - \epsilon_{\m{Q}} - \Sigma(\omega, \m{Q})},
  \label{G2}
\ee
where the ``self-energy'' $\Sigma$ usually has both real and imaginary parts. The former describes the spectrum renormalization, whereas the latter is related to the magnon lifetime. If magnons are well-defined quasiparticles (which is usually the case in magnetically ordered phases) then one can use the ``on-shell'' approximation $\Sigma(\omega, \m{Q}) \rightarrow \Sigma(\epsilon_{\m{Q}}, \m{Q})$. Therefore,  instead of delta-function shape (see Eq.~\eqref{G1}) the magnon spectral weight $-\textrm{Im} \, G(\omega, \m{Q})/\pi$ acquires Lorentzian form:
\be
  -\frac{\textrm{Im} \, G(\omega, \m{Q})}{\pi}= \delta(\omega - \epsilon_{\m{Q}}) \rightarrow \frac{1}{\pi} \frac{\Gamma_{\m Q}}{(\omega - \epsilon_{\m{Q}})^2 + \Gamma^2_{\m Q}}.
\ee
The damping rate $\Gamma_{\m Q}$ is inversely proportional to magnon lifetime. 

When addressing the scattering cross-section (Eq.~\eqref{A1}) its intensity is proportional to the corresponding imaginary part of the spin susceptibility ~\cite{maleev2002polarized}. So, in order to account for the finite magnon lifetime we modify Eq.~\eqref{A1} by the substitutions:
\be
  \delta(\omega-\epsilon_{\m Q}) \rightarrow \frac{1}{\pi} \frac{\Gamma_{\m Q}}{(\omega - \epsilon_{\m{Q}})^2 + \Gamma^2_{\m Q}}, \nonumber \\
  \delta(\omega +\epsilon_{-\m Q}) \rightarrow \frac{1}{\pi} \frac{\Gamma_{-\m{Q}}}{(\omega + \epsilon_{-\m{Q}})^2 + \Gamma^2_{-\m Q}}.
  \label{subs}
\ee
One should also make these substitutions in Eq.~\eqref{Int1}. After that, in principle, the analytical integration is still possible, however the result is very cumbersome and can barely be used in practice. Instead, we take advantage of the Lorentzian approximation and the fact that the main signal comes from the same frequencies as in the case of delta-functions considered before. Assuming that the other factors in the integral~\eqref{Int1} are slowly varying functions in comparison with the magnon spectral weight, we can take them away from the integral (this assumption is valid in vicinity of $\t_C(H)$ and at larger angles). Then, for contributions at $\t_x>0$ we have the integral
\be
  \int dt \, \textrm{Im} \frac{1}{ - \frac{E_i}{\t_0} \left[ (t-\t_0)^2 + \t^2_{rel} -\t^2_C \right] + i \Gamma_{\m Q}} \propto \nonumber \\
  \int dt \, \textrm{Im} \frac{1}{ (t-\t_0)^2 + \t^2_{rel} -\t^2_C - i \gamma_{0}}, \label{Int2}
\ee
where $\m{Q}$ is determined by the energy and momentum conservation laws, and $\gamma = \t_0 \Gamma/E_i $, which in our simplified calculations can be taken as $\gamma_0$ which corresponds to $t = \t_0$ and magnon momentum $\m{Q} = k_i (\t_B+\t_C(H),0,\t_0)$ (or to any momentum with the same $\t_{rel}=t_C(H)$). Now, we can express the imaginary part of the integral~\eqref{Int2} as
\be
  \sigma_{ch}(\t_{rel}) \propto \textrm{Im} \frac{1}{\sqrt{\t^2_{rel} -\t^2_C - i \gamma_{0}}}
  \label{Asigma2}
\ee
which is an extension of Eq.~\eqref{Asigma1} onto the case where magnon damping is included. One can see that now the scattering intensity is nonzero also in $\t_{rel} > t_C(H)$ region, which is the effect of finite magnon lifetime. Another important feature of Eq.~\eqref{Asigma2} is the finite signal $\propto 1/\sqrt{\gamma_{0}}$ at the cutoff ($\t_{rel} = \t_C(H)$). It thus becomes clear that introducing magnon damping leads to a smoothing of the SANS intensity near the cutoff, and a power-law decaying tail at $\t_{rel} > \t_C(H)$.

SANS spectra where magnon damping is taken into account, are shown in Figs.~\ref{FigApp}(b) and (c). Dashed lines therein are simplified spectra obtained with the use of Eq.~\eqref{Asigma2}, they are not the best fits of corresponding ``exact'' solid curves. We just multiply them by a factor $60$ in order to compare two ways of calculating the spectra. Evidently, the approximate calculations cannot describe the scattering intensity near $t_B$ (small $\t_{rel}$) at small gaps $\Delta$ ($\t_C \approx \t_0$). Nevertheless, if the intensity curve has a local maximum (lower damping), then this data can be approximated using the data in the vicinity of this maximum and larger angles. In this case, the $\t_C$ parameter approximately corresponds to the maximum intensity angle. In the opposite case of larger damping, when the local maximum is absent, the inflection point (minimum of the intensity derivative with respect to the angle) in the large $\t_{rel}$ region should play the same role. One can also see in Fig.~\ref{FigApp}(c) that the smaller $\t_C$ is, the larger is the discrepancy between numerical integration result and the one obtained using Eq.~\eqref{Asigma2}. The reason is that at $\gamma_0 \gtrsim \t^2_C$ Eq.~\eqref{Asigma2} the assumption of sharp magnon spectral weight in the background of other slowly-varying factors (see Eq.~\eqref{Int1}) breaks down. Moreover, in this regime the magnon lifetime is short, and the resulting SWSANS spectrum looks like diffuse scattering rather than inelastic scattering on well-defined quasiparticles. In this case, the cutoff angle does not have a clear physical meaning.

\end{document}